\documentclass[12pt, a4paper]{article}
\usepackage{setspace}
\usepackage[utf8]{inputenc}
\usepackage[english]{babel}
\usepackage{amsmath,amsfonts,amssymb,amsthm}
\usepackage{bm}
\usepackage{graphicx}
\usepackage{url}
\usepackage[dvipsnames]{xcolor}
\usepackage{layout}
\usepackage{subcaption}
\usepackage{listings}
\usepackage{xcolor}
\usepackage{tabularx}
\usepackage{subcaption}
\usepackage{changepage}
\usepackage{array}
\usepackage{caption}
\usepackage{float}
\usepackage{fancyhdr}
\usepackage{adjustbox}
\usepackage{graphicx}
\usepackage{authblk} 
\usepackage{adjustbox}
\usepackage{multicol}

\usepackage{geometry}
\title{Environmental monitoring using orbital angular momentum mode decomposition enhanced machine learning}

\author[1]{Zhaozhong Chen}
\author[1]{Ultan Daly}
\author[1]{Aleksandr Boldin}
\author[1]{Lenny Hirsch}
\author[1,2]{Mingjian Cheng}
\author[1]{Martin P.J. Lavery \thanks{Corresponding author: Martin.Lavery@glasgow.ac.uk}}

\affil[1]{James Watt School of Engineering, University of Glasgow, Glasgow, UK, G12 8QQ}
\affil[2]{School of Physics, Xidian University, Xi’an 710071, China}

\begin{document}
\maketitle

\begin{abstract}
Atmospheric interaction with light has been an area of fascination for many researchers over the last century. Environmental conditions, such as temperature and wind speed, heavily influence the complex and rapidly varying optical distortions propagating optical fields experience. The continuous random phase fluctuations commonly make deciphering the exact origins of specific optical aberrations challenging. The generation of eddies is a major contributor to atmospheric turbulence, similar in geometric structure to optical vortices that sit at the centre of OAM beams. Decomposing the received optical fields into OAM provides a unique spatial similarity that can be used to analyse turbulent channels. In this work, we present a novel mode decomposition assisted machine learning approach that reveals trainable features in the distortions of vortex beams that allow for effective environmental monitoring. This novel technique can be used reliably with Support Vector Machine regression models to measure temperature variations of $0.49$\,$^\circ$C and wind speed variations of $0.029\,$m\,s$^{-1}$ over a 36\,m experimental turbulent free-space channels with controllable and verifiable temperature and wind speed with short 3\,s measurement. The predictable nature of these findings could indicate the presence of an underlying physical relationship between environmental conditions that lead to specific eddy formation and the OAM spiral spectra.
\end{abstract}

\newpage
\maketitle

\section{Introduction}
The interaction of optical fields with disordered and complex media has been an area of considerable interest for many researchers \cite{FRIED:1965gu,Mosk2012,Ladyman2013WhatSystem,Kajenski1992,Cizmar2012ExploitingImaging,Caramazza2019TransmissionFibre,Valencia2020UnscramblingMedium}. Such media commonly induces intensity and phase distortions that can be disruptive for optical communications and imaging systems \cite{cao2020distribution,Huang:2015ht,Lavery2017Free-spaceEnvironment,Paterson:2005wj,Trichili2020RoadmapOptics,Ren:2013ht,Babcock1953Oct,Hampson2021}. However, these optical distortions are directly dependent on the environment that the complex media is situated within and therefore measuring these distortions could be used as a novel way to sense changes in the world around us \cite{FRIED:1965gu,Trichili2020RoadmapOptics,Lavery2017Free-spaceEnvironment,Viola2020DegradationScattering}. Free-space optical applications, such as space division multiplexing, quantum key distribution, optical imaging, and optical sensor systems are all acutely affected by the atmosphere \cite{Andrews2019}. The core challenge faced during propagation through the atmosphere is that these optical beams experience path dependent optical distortions that are accrued over hundreds or thousands of meters of propagation \cite{Primmerman1995}. Along this path, there is time-dependent and random variations in the temperature and wind speed of the air resulting in density changes  \cite{Strohbehn1968}. These subsequently result in spatial and temporal changes in the refractive index structure of the propagation channel that lead to complex optical aberrations that can disrupt imaging, sensing, and communication systems \cite{Hampson2021,Sathe2015,Zhu2002}. One burgeoning area of research is mitigation of the crosstalk from the aberrations that arise in free space optical (FSO) channels to enable space division multiplexing (SDM), where turbulence leads to crosstalk between independent data channels and can dramatically limit the performance of communication systems \cite{Li2023Jul}. It is common to describe each channel in an optical SDM system as a set of orthogonal spatial modes such as Laguerre Gaussian or Hermite Gaussian modes. Laguerre-Gaussian (LG) modes are naturally eigenmodes of circular apertures, where these modes are characterized by orthogonal mode indices $\ell$ and $p$, corresponding to the azimuthal and radial components, respectively \cite{Abramochkin1991}. These beams have a transverse complex amplitude profile of $A\left(r\right) \exp\left(i\ell \phi \right)$ and carry an Orbital Angular Momentum (OAM) of $\ell \hbar$ per photon with $r$ and $\phi$ as the radial and angular coordinates, respectively \cite{Allen:1992vk}. Optical aberrations or misalignment between the propagation axis of the optical field and axis of the receiver have been shown to have distinctive crosstalk between received modes that has been referred to as OAM spectroscopy \cite{Paterson:2005wj,Vasnetsov2005,Lavery2011-Spectra}. These spectra have previously been used to realise digital spiral imaging for determining changes of phase and amplitude of an optical beam \cite{Molina-Terriza2001,Torner2005,Molina-Terriza2007}. During propagation in long-distance turbulent channels, each LG mode experiences unique optical distortions \cite{Lavery2017Free-spaceEnvironment,Lavery2018VortexPropagation}. These OAM dependent distinctive interactions could be leveraged to extract the environmental origins of the observed inter-channel crosstalk and used to create enhanced hybrid communication and environmental sensing systems.

Turbulence in the atmosphere can be generated by the mixing of air with different temperatures, where this mixing forms eddies as the energy is dissipated. These eddies have a spiral structure, similar to that of hurricanes, where their size, distribution, and movement are a direct result of weather parameters like temperature distributions and wind speed \cite{cheng2021}. The commonly observed optical aberrations in free-space are a direct result of the optical complex amplitude with the spatial variation in density from propagation through many optical eddies over a given channel. Each interaction will result in local changes in the direction of propagation and phase, which after propagation will lead to spatial changes in intensity that evolve during propagation as interference occurs between the locally tilted wavefronts. This effect is the key contributor to scintillation, where experimental techniques such as scintillometry track the time varying intensity changes and computationally compare them to estimated channel models to infer environmental parameters \cite{Andreas1988}. Typically, commercial scintillometers measure the refractive index structure parameter, $C_\mathrm{n}^2$, which can give an estimate of heat variation across an operational range that can span from hundreds of meters to several kilometers \cite{Andreas1988,Hill1997}. These measurement approaches typically require long integration times (ten’s of minutes to over one hour) to achieve acceptable accuracies for particular applications. 

Solely measuring global changes in intensity does not allow for direct determination of the key parameters such as inner scale value, $l_\mathrm{0}$, which are closely related to the structure of the eddies in turbulence. Multi-aperture scintillometers (MASS) have been suggested to improve the resolvability of $l_\mathrm{0}$ and $C_\mathrm{n}^2$, where multiple apertures simultaneously record the intensity variations \cite{Hill1997,Kornilov2007CombinedStudies}. Recently, wavefront sensors combined with machine learning have been demonstrated to offer further improvement by substantially increasing the number of effective apertures \cite{He2021May}. However, the natural evolution of eddies has a rotational spatial structure, therefore measurements of the spiral motion will be projected into a pixelated coordinate system. Conversely, OAM beams twist as they propagate creating an optical vortex along their axis of propagation, which is structurally similar to atmospheric eddies created by turbulence. This suggests measurements of OAM content from light interaction with atmospheric eddies could provide improved determination of key parameters about a particular physical channel.

\begin{figure}[H]
   \center
   \includegraphics[width=0.75\textwidth]{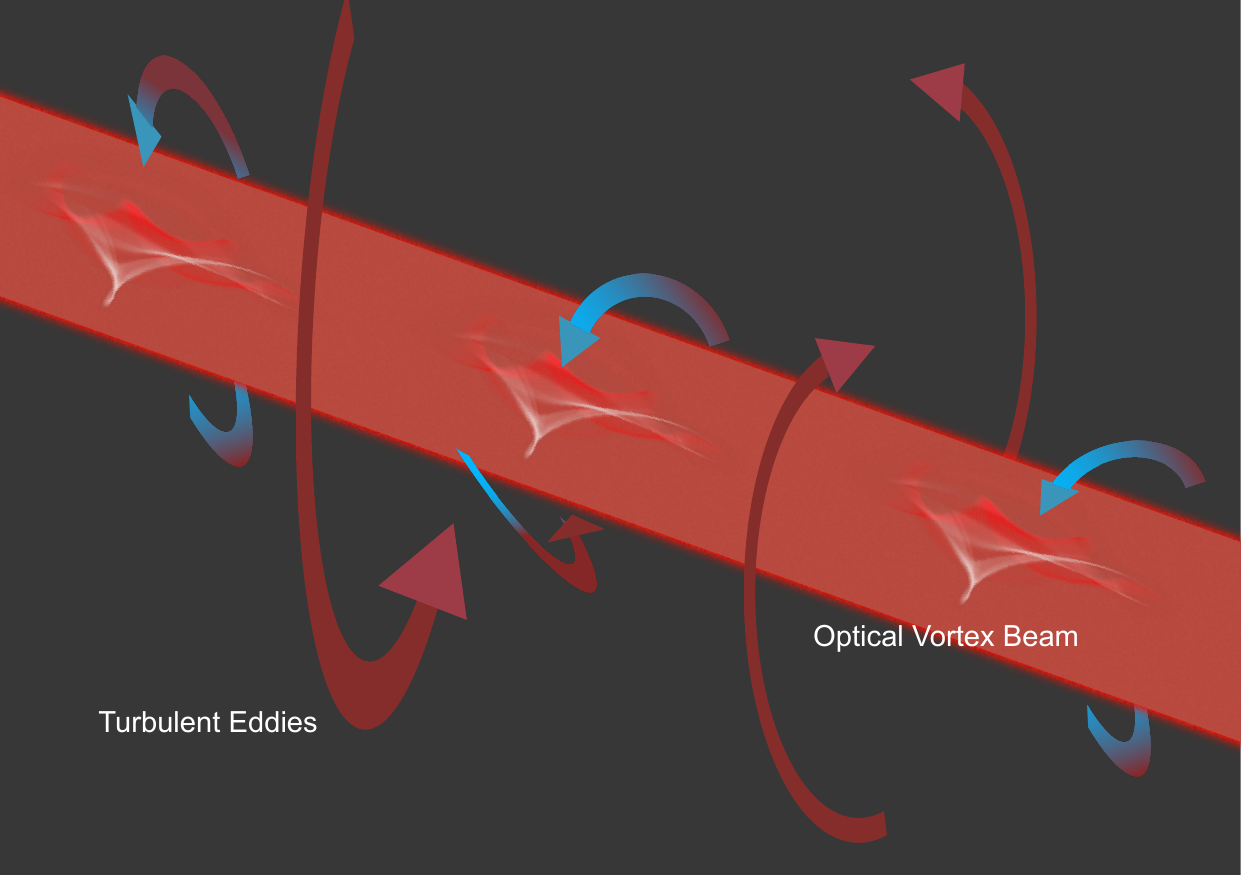}
   \caption{Concept. Turbulence eddies interact uniquely with twisted optical beams such as those that carry OAM.}
   \label{fig:Concept}
\end{figure}

This article demonstrates experimentally that the use of an optical probe carrying OAM and an OAM sensitive receiver can be used as a tool to measure both the wind speed and temperature in an electronically controlled 36\,m turbulent FSO channel. A measurement resolution of 0.49$^{\circ}$C and wind speed variations of $0.029$\,ms$^{-1}$ were achieved via the use of a Support Vector Machine (SVM) regression algorithm, trained on the inter-channel crosstalk between OAM modes measured by a passive mode sorter, using an $\ell=3$ optical probe beam. The OAM mode-sorting can be considered as a form of all-optical pre-processing to support feature discovery and experimentally revealed features that were embedded in the OAM spectra that could be used reliably with machine learning regression models. The recorded crosstalk gives a measurement of the power distribution of OAM within an optical beam concerning a specific measurement axis. Changes in the total OAM or movement in the position of an optical vortex will lead to distinctive power distributions, also referred to as OAM spectrum, that can be used for monitoring the structure of turbulent fields. When a Gaussian mode that has OAM of $\ell=0$ propagates through the turbulence induced atmospheric eddies, randomly distributed pairs of phase vortices are created with $\ell= -1$ and $\ell=1$. As pairs are produced, the total OAM $\ell=0$ is maintained due to conservation of angular momentum \cite{HECKENBERG:1992vn, Lavery2018VortexPropagation, Lavery2017Free-spaceEnvironment}. However, the center of each vortex will not align with the measurement axis and therefore results in the distribution of OAM with a mean not equal to the $\ell$ of the input beam and with a width determined by the number and distribution of vortices. As the channel distorts this beam, the variation in the mean OAM will be directly related to the interactions of turbulence and the light mode \cite{Lavery2018VortexPropagation,Lavery2017Free-spaceEnvironment}. Therefore, measurement parameters such as average OAM and OAM distribution are closely related to the environmental conditions of the channel. We expect this technique could be used as an effective path integrated and potentially channel resolved. This approach could also be used as a  weather monitor to increase the resolution of weather prediction systems \cite{Klaver2020,Pielke1991,LINDBORG1999}. This is a key challenge in predicting extreme weather events, partly due to the limited sampling generally available for weather mapping from highly localised weather stations \cite{WMO2023}. Path-integrated weather monitors could provide unique data for future weather models as this approach can provide an increased accuracy over scintillation index based measurements of wind-speed and temperature measurements by a factor of 4.49 and 3.55, respectively. Further, we believe our technique could be used for tough environments such as jet engine thermal exhaust monitoring. The system architecture is also ideal for hybrid FSO communications and environmental sensing as the de-multiplexer could be used for both SDM channel demultiplexing and challenge monitoring simultaneously.

\section{Atmospheric Turbulence}
Atmospheric turbulence is governed by the fluid dynamics of a particular channel and has a key parameter called Reynolds number, $Re$. This number is a numerical relationship between viscous and inertial forces, where low $Re$ signifies the dominance of viscous forces leading to constant flow and high $Re$ signifies inertial forces dominance leading to more chaotic behaviour. 
This can be expressed as $Re = \frac{VL}{\nu}$, where $\nu$ is the kinematic viscosity, $V$ is the velocity of flow, and $L$ is the characteristic linear dimension of the fluid, respectively \cite{Richardson1922}. Fluid flows can be categorised into two main types: laminar and turbulent flows. Laminar flow is a smooth, orderly movement of fluid with parallel layers and minimal mixing, typical of slow-moving, viscous fluids. In turbulent flow, the velocity and direction of flow at a specific point undergo continuous and random changes, resulting in eddies, and rapid fluctuations in flow characteristics. In 1922, Lewis Fry Richardson developed one such early model known as the energy cascade theory, which explains turbulence as a cascade of random turbulent eddies with various scale sizes, that range from the smallest eddy known as the inner scale $l_\mathrm{0}$ to an outer scale, $l_\mathrm{0}$ that defines the largest thermally stable volume of air \cite{Richardson1922}. Large areas of air are heated, by the sun for example, forming large scale turbulent eddies, $L_\mathrm{0}$, which subsequently break down further into smaller eddies under the influence of inertial forces until the eddy reaches the inner scale when the energy within that eddy dissipates into heat. For the majority of naturally occurring turbulent environments, we only consider the propagation of optical beams through a restricted range of Reynolds numbers linked to the ratio of inner scale to outer scale by the relationship, 

\begin{equation}
Re = \left(\frac{L_\mathrm{0}}{l_\mathrm{0}}\right)^{\frac{4}{3}}.    
\end{equation}

\noindent The critical parameter that affects optical fields are the refractive index fluctuations that result from the eddies formed \cite{Richardson1922}. The turbulence power spectrum models can give the statistical averages of the random variation process of the atmosphere. The turbulence mixes different layers of air forming eddies and as pressure is locally constant, this results in a spatial variation of density, $\rho$, that leads spatial variation in refractive index, $n$.  The average size of the smallest turbulent eddies,

\begin{equation}
    l_\mathrm{0}=\left (\frac{\nu^3}{\epsilon} \right)^{\frac{1}{4}} \label{eq:l0}
\end{equation}

\noindent where $\nu$ is the kinematic viscosity and $\epsilon$ is the energy dissipation rate of atmosphere. Kinematic viscosity $\nu$ is a function of temperature and $\epsilon$ is related to both temperature and wind speed distribution \cite{kolmogorov1941local}. The refractive-index structure constant $C_\mathrm{n}^2$ can be deduced from the temperature structure constant $C_T^2$ by

\begin{equation}
      C_\mathrm{n}^2 =\left(79 \times 10^{-6} \frac{P}{T^2} \right)C_T^2 \label{eq:cn2}
\end{equation}

\noindent where $P$ is pressure, $T$ is temperature, and $C_T^2$ is the temperature structure constant.  $C_\mathrm{n}^2$ is a measure of the local turbulence strength. The phase distortion can be described by a coherence length scale of the turbulence known as the Fried parameter $r_\mathrm{0}$. 

\begin{equation}
     r_\mathrm{0}=\left[0.42 \sec(\zeta k^2 \int^l_\mathrm{0}  C_\mathrm{n}^2(z)\mathrm{d}z) \right]^{-\frac{3}{5}}. \label{eq:r0}
\end{equation}

\noindent Here, $ C_\mathrm{n}^2$ is the refractive index structure constant, $k$ is the wave number, and $\zeta$ is the zenith angle \cite{FRIED:1965gu}. Through the above equations and analysis, it is expected that both temperature and wind speed have a joint effect on the turbulence parameter $r_\mathrm{0}$ and $l_\mathrm{0}$. 

\begin{figure}
   \center
   \includegraphics[width=0.82\textwidth]{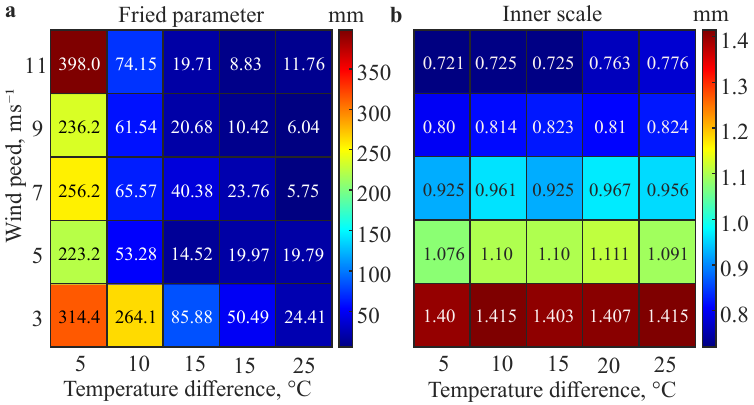}
   \caption{Turbulence parameters from CFD. \textbf{a}, the Fried parameters under the same 25 different temperature and wind speed combinations. \textbf{b}, the inner scale under 25 different temperature and wind speed combinations. }
   \label{fig:l0r0}
\end{figure}

\begin{figure}
   \center
   \includegraphics[width=0.85\textwidth]{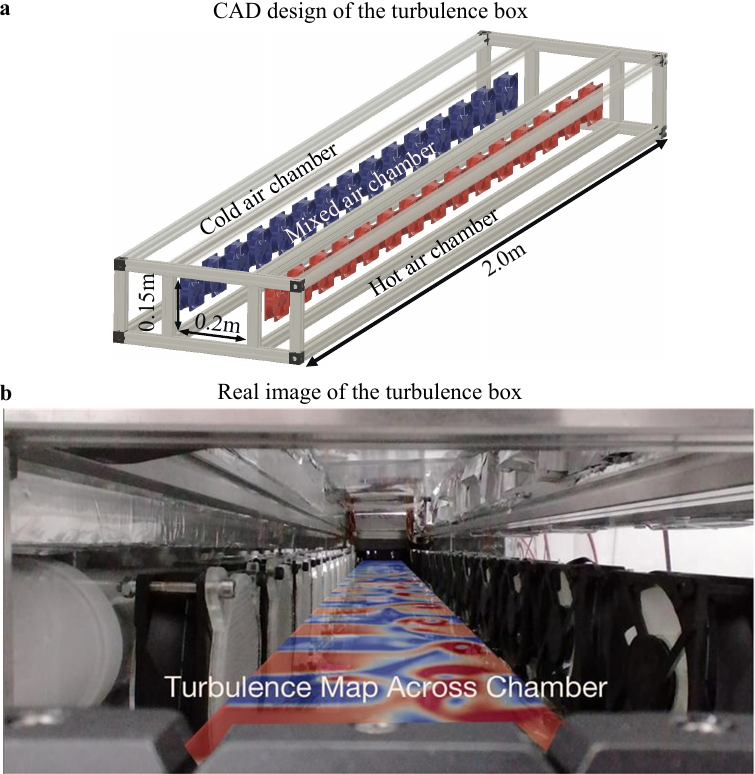}
   \caption{Turbulence box design. \textbf{a}, The CAD design of the box. There are cold, hot, and mixed chambers. The dimensions of mixed chambers are $0.15$\,m, $0.2$\,m, and $2$\,m in height, width, and length, respectively. \textbf{b}, The actual image of the turbulence box is superimposed onto the temperature distribution figure obtained from CFD.}
   \label{fig:Boxdesign}
\end{figure}

Turbulent flow is a nonlinear and multiscale process governed by the Navier-Stokes equations. As direct solving of Navier-Stokes equations are computationally challenging, therefore statistical models are commonly used to represent the interactions in atmospheric channels instead of purely computational approaches. One commonly used model is Large Eddy Simulation (LES), which aims to reduce computational costs by filtering out the smallest length scales, which are computationally intensive, through low-pass filtering of the Navier–Stokes equations, effectively eliminating fine scale information \cite{Deardorff1970Apr}. Using computational fluid dynamics (CFD) modeling software (OpenFOAM v2212), we implemented an LES to create a turbulence channel numerically that could provide an estimate of the spatial and temporal variation of key turbulence parameters. This was subsequently used to optimize the design of an experimental turbulence simulator. Turbulence is produced in a laboratory setting by mechanically forcing the mixing between warm and cooled air in a central mixing area \cite{Keskin2006Jul}. To replicate this, we simulated a system that includes heat reservoirs for both warm air and cool air that will be mixed in a central chamber, where Fig. \ref{fig:Boxdesign}a shows the CAD design of the atmosphere turbulence simulator. The CFD mesh was created to represent a physical turbulence chamber that is $2\,\text{m}$ in length, $0.15\,\text{m}$ in height, and $0.2\,\text{m}$ wide with 15 air inlets for each heat reservoir. To relate the fluid dynamic properties of the channel to turbulence parameters $l_\mathrm{0}$ and $r_\mathrm{0}$, in the CFD simulation spatial field distribution of $T_\mathrm{cfd}$, $P_\mathrm{cfd}$, and $\vec{V}_\mathrm{cfd}$ was recorded every millisecond for the mixing chamber and calculated by equation \eqref{eq:l0} and \eqref{eq:r0} for $l_\mathrm{0}$ and $r_\mathrm{0}$, respectively. To investigate a range of different environmental conditions the speed of the air from the reservoir to the mixing chamber, $V$, was varied, along with different temperature differences across the chamber. Where, $V$, was varied from 3\,m\,s$^{-1}$  to 11m\,s$^{-1}$  in intervals of 2\,ms$^{-1}$ and the temperature difference between hot air $T_\mathrm{hot} = T_\mathrm{cool} + \Delta\,T$ and $\Delta\,T$ ranges from 5$^\circ$C to 25$^\circ$C in intervals of 5$^\circ$C,  $T_\mathrm{cool}$ was fixed at 21 $^\circ$C. The calculated  $l_\mathrm{0}$ and $r_\mathrm{0}$ are shown in Fig. \ref{fig:l0r0}, where a distinctive relationship between these values of temperature and flow rate is observed. 

\section{Structured Light Interaction with Environment}
LG modes form a complete basis set of orthogonal modes, where any complex amplitude profile can be represented as a weighted superposition of these spatial modes, known as a spatial modal spectrum. A notable behaviour observed for OAM modes that are aberrated is called vortex splitting, where a beam that carries an OAM  $\ell=m$, will break up into $m$ vortexes of $\ell= 1$ that are spatially separated. The total orbital momentum is conserved but unlike in an undisturbed mode, it is not entirely localized at the original beam propagation axis and where the average distance of movement away from the beam axis is directly related to the $D/r_\mathrm{0}$ of the atmospheric channel \cite{Lavery2018VortexPropagation}. To obtain the expected OAM spectrum after propagation through a bulk turbulent channel with different combinations of $r_\mathrm{0}$ and $l_\mathrm{0}$, we performed a propagation simulation based on the split-step algorithm using 120 segments to simulate a representative bulk turbulence channel. Each screen was calculated using the von Kármán spectrum and implementing sub-harmonic correction \cite{McGlamery1976Jul, Lavery2018VortexPropagation}.

\begin{figure}
   \center\includegraphics[width=0.80\textwidth]{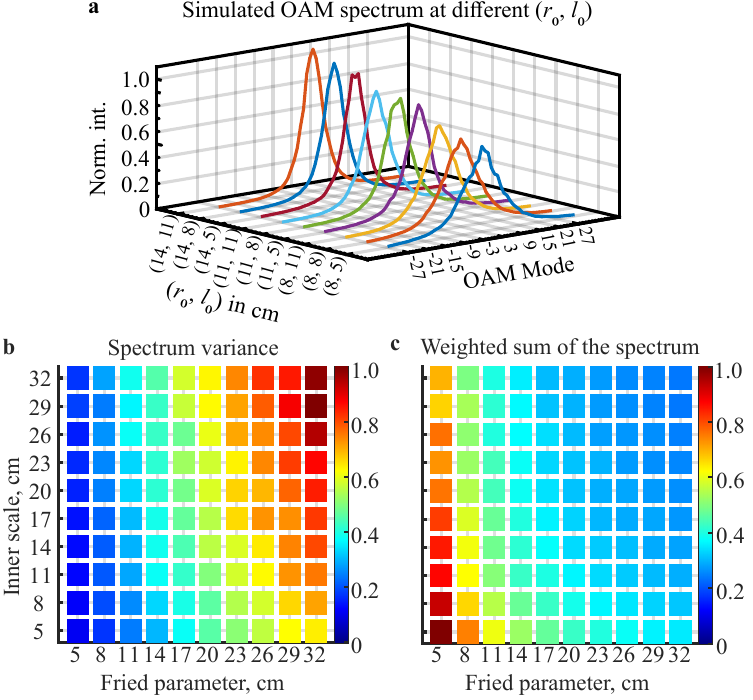}
   \caption{ Simulated OAM spectrum and normalized features. \textbf{a}, OAM spectra for a range of $l_\mathrm{0}$ and $r_\mathrm{0}$ values when an OAM beam with $\ell=3$ is propagated over a simulated optical channel. \textbf{b}, The spectrum variance, $\sigma^2$, is calculated at 100 different different turbulence channels with split-step propagation simulation. \textbf{c}, The weighted sum of the spectral components $\eta$, calculated at 100 different turbulence channels with split-step propagation simulation.}
   \label{fig:simulatedspectrum}
\end{figure}

To investigate the specific effect on the OAM spectrum through the simulated turbulent channel, 10 Fried parameters, $r_\mathrm{0}$, and 10 inner scales, $l_\mathrm{0}$ were tested, where each combination was simulated to provide 100 effective channel conditions. For each channel, 100 realizations were computed to account for the random variation in the channel. The spectra for a beam with $\ell=3$ and $p=0$ with selected spectra are shown in Fig.\ref{fig:simulatedspectrum}a. The spectra features were subsequently analysed by considering a weighted sum of the spectral components, 
\begin{equation} \label{eq:n}
    \eta = \sum_{\ell_\mathrm{min}}^{\ell_\mathrm{max}}A_\ell^2 \cdot |\ell|,
\end{equation}
\noindent and the variance in the spectra,
\begin{equation}  \label{eq:sig2}
     \sigma^2 = \frac{1}{N} \sum_{\ell_\mathrm{min}}^{\ell_\mathrm{max}}(A_\ell^2 - \mu)^2 ,
\end{equation}
\noindent where $A_\ell$ is the normalised amplitude of each OAM mode with a spiral phase index of $\ell$, $\mu = \frac{1}{N} \sum_{\ell_\mathrm{min}}^{\ell_\mathrm{max}}A_\ell^2$ and $N$ is the total number of OAM mode. Changes in $r_\mathrm{0}$ and $l_\mathrm{0}$ lead to continuous variation both $\eta$ and $ \sigma^2$, as shown in Fig.\ref{fig:simulatedspectrum}b,c. In real air turbulence channels, $r_\mathrm{0}$ and $l_\mathrm{0}$ are functions of $T$ and $V$. Therefore, we propose that each can be defined as a function $\eta = f(r_\mathrm{0}, l_\mathrm{0}) = g(T, V)$ and $\sigma^2 = f'(r_\mathrm{0}, l_\mathrm{0}) = g'(T, V)$. Explicit derivation of these functions is challenging due to the statistical approaches to representing real world turbulence accurately. However, regression analysis of real world turbulence can provide a fitted model that can strongly indicate the existence of a direct connection between the OAM spectrum and changes in $T$ and $V$.

\section{Results}
\begin{figure}
   \center
\includegraphics[width=0.92\textwidth]{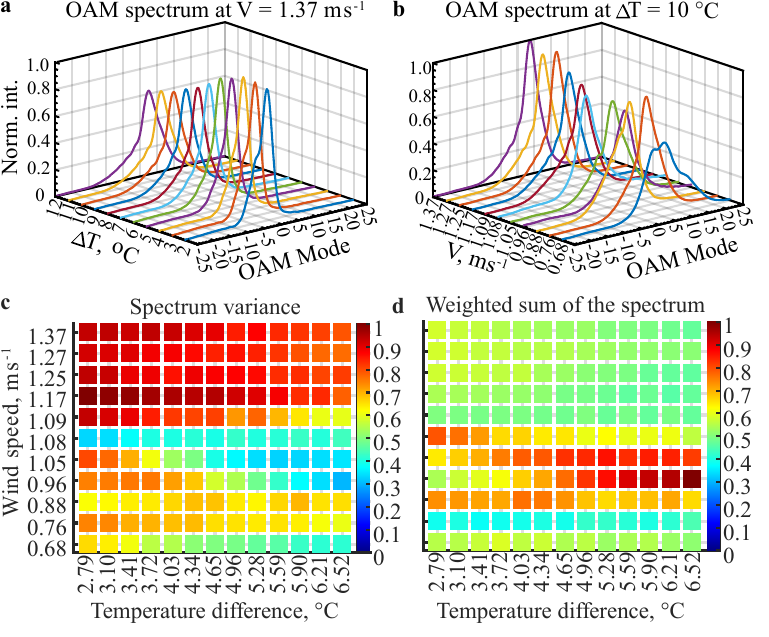}
   \caption{ Experimental OAM spectrum for $l=3$ at different wind speeds and temperatures. \textbf{a} OAM spectrum curves when the wind speed is V = 1.37 m\,s$^{-1}$ and the $\Delta\,T$ vary in the range of $2$ to $12^\circ\text{C}$. \textbf{b}, OAM spectrum curves when $\Delta\,T = 10^\circ\text{C}$ and the wind speed vary in the range of $0.68$ to 1.37\,m\,s$^{-1}$. $\eta$ and $\sigma^2$. \textbf{c}, The experimental spectrum variance, $\sigma^2$. \textbf{d}, The weighted sum of the experimental spectral components $\eta$.}
   \label{fig:experimentalspectrum}
\end{figure}

Based on the numerical simulations, an experimental environmental simulator was developed to emulate the multiple interactions with turbulence expected in long distance optical propagation. This was achieved with a folded mirror arrangement, which was used to achieve a total propagation length of 36\,m. Each reflection is offset, to experience a different effective turbulence induced phase aberration (Section \ref{sec:Methods}, methods). The simulator was heated initially to have an air temperature around $15^{\circ}$\,C above ambient. While continually monitoring temperature and windspeed, the temperature difference and average windspeed were varied to capture 1012 unique environmental conditions. For each condition, optical beams with between $\ell=-3$ and $\ell=3$ were transmitted over the turbulence channel. To analyse the OAM spectrum, a passive mode sorter was used to decompose the incoming beam into 51 equal sized regions on a CCD camera. The pixels in each region were summed and are used to measure the weighting of the different OAM spectral components. OAM spectrum at a frame rate of 30\,Hz and the spectra were recorded for each condition. The Greenwood frequency of the channel varies with change in windspeed and the turbulence strength. The maximum measured frequency was approximately 388\,Hz, corresponding to a coherence time of 2.58\,ms. All data was labeled with time stamps, real-time temperatures, and wind speed to allow for accurate data analysis and machine learning training. The variation of the OAM spectra can be seen to have a distinctive structure that varies with respect to the changes in temperature, Fig.\ref{fig:experimentalspectrum}a, and changes in windspeed, Fig.\ref{fig:experimentalspectrum}b. For each condition, the 3\,s of recorded spectra are summed to reduce the effect from random spectral features that arise from complex turbulence processes  \cite{Lavery2017Free-spaceEnvironment}. The obvious similarity between the adjacent curves in Fig.\ref{fig:experimentalspectrum} \textbf{a},\textbf{b} indicates that the OAM spectrum is not entirely random. These similar features serve as the basis for training our SVM model.

To analyse the spectra, we can compute two metrics for the experimental observations, $\eta$, equation (\ref{eq:n}), and $\sigma^2$, equation (\ref{eq:sig2}). Both metrics are have distinctive structure, suggesting an underlying function determining the observed result. It is proposed that the intersection between these underlying functions can be used to simultaneously measure the windspeed and temperature of our turbulent channel simulator. However, due to the coupled nature of the system, machine learning tools can be readily used to determine a hyperplane defined by input features. Therefore, an SVM algorithm is used to perform regression analysis. The kernel function is a critical component that determines the shape and dimensionality of the decision boundary. We chose a polynomial kernel to map the data into a higher-dimensional space using a polynomial function.

\begin{equation}
     K(X_i,X_j) = (\alpha\,X_i\cdot\,X_j + \beta)^\gamma
\end{equation}

\noindent where, $\alpha$, $\beta$, and $\gamma$ are kernel parameters that influence how the data is transformed into a higher-dimensional space. $X_i$ and $X_j$ represent the feature vectors. The parameter $\alpha$ governs the impact of individual training samples on the decision boundary. Higher values of $\alpha$ result in more elaborate decision boundaries that closely align with the training data. On the other hand, $\beta$ introduces an offset to the decision boundary, while $\gamma$ denotes the polynomial degree within the kernel function, thereby dictating the complexity of the decision boundary.

\begin{figure}
   \center
   \includegraphics[width=0.95\textwidth]{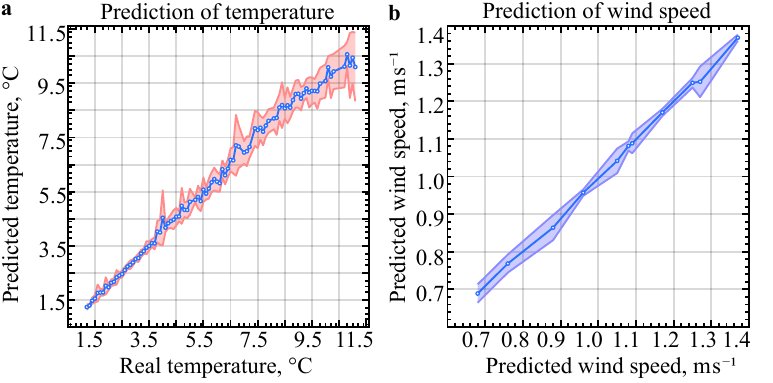}
   \caption{Predictions of temperature and wind speed with the trained SVM model. \textbf{a}, Prediction of temperature difference. \textbf{b}, The prediction of temperature difference.}
   \label{fig:SVMPrediction}
\end{figure}

The OAM spectrum with range $\ell=\pm25$ was used as the input for SVM training, where the data was labelled for both windspeed and temperature difference from the channel emulator. These labels were determined by measuring the average of 10 thermometers spaced evenly over the channel emulator and calibrated tachometers on each fan. The polynomial order of $\gamma = 2$ is chosen to prevent overfitting of the data, due to random fluctuations arising from the turbulence. Each training data point is the average of 90 spectra, which represents the data collected over 3 seconds. In total, we utilise 3,897 training data points to compute our prediction model. Our data was divided into training and validation groups randomly across multiple realizations of the channel, where a 5-fold cross-validation was implemented. In total, 3 groups of data collected on different days were aggregated to ensure results are not solely related to a specific channel realization. The prediction of wind speed and temperature of the trained SVM model are shown in Fig. \ref{fig:SVMPrediction} \textbf{a} and \textbf{b}, respectively. We trained two separate models using the same training data to predict wind speed and temperature, and we used the standard deviation of the prediction results to represent the accuracy of the predictions. Th highest accuracy achieved for temperature and windspeed were predicted to be $0.49^\circ$C and $0.029$\,m\,s$^{-1}$ respectively.

\section{Discussion and conclusions}
Our experimental results have focused on the decomposition of the optical field into its constituent OAM modes. However, other methods for analysing the received optical intensity can be considered. To determine the benefit of OAM decomposition, training based on the received intensity profile and scintillation index, a commonly used metric for analysing turbulent behaviour, were considered. Using direct imaging onto a CCD camera, the optical field as it passed through the atmospheric channel, and the intensity profile were measured simultaneously with the OAM decomposed output to allow for direct comparison with the same environmental conditions. The SVM model was trained using two vectors, which were the summation of the pixels along horizontal and vertical, respectively averaged over 3s of data. Again, using the data from the directly imaged optical field, the scintillation index was calculated over a 3\,s time window, ensuring that the size of training data was the same as that used for the SVM models based on OAM spectrum and intensity profile measurement. Therefore, in the presented results the scintillation index for each 302 pixels that the beam covered on the camera was determined and used for model training. This is equivalent to dividing the receiver into 302 independent apertures. The direct comparison of these results is indicated in Table \ref{tab:Comparison}. 

Comparing measurements with intensity profile and OAM spectrum, we find similar levels of accuracy. However, when utilizing an OAM probe $|\ell|=3$ for both temperature and wind speed measurements, we observe an average improvement factor of 1.21.
The difference in performance for OAM spectrum measurements is more noticeable for wind-speed measurements with an average accuracy gain factor of 1.37. A gain factor of 1.33 was measured using $|\ell|=3$ OAM modes for wind-speed measurements. As each specific channel is slightly different due to random mixing, for wind speed and temperature, where coupled accuracy variation can be seen across all three measurement types. 

These findings show that analyzing intensity profile or OAM spectrum yields significantly better accuracy than analyzing scintillation with an improvement factor of over 3 observed in temperature and wind speed measurements. In addition, these results suggest that using the OAM spectrum probe leads to  improvements over the scintillation index in both temperature and wind speed measurements, with improvement factors of 3.49 and 4.25, respectively.

\newcolumntype{C}[1]{>{\centering\arraybackslash}p{#1}}

\begin{table}[H]
\centering
\begin{tabular}{|C{1.2cm}|C{2cm}|C{2cm}|C{2cm}|C{2cm}|C{2cm}|C{2cm}|}
\hline
 & \multicolumn{3}{c|}{$\Delta T (^\circ C)$ prediction-RMSE} & \multicolumn{3}{c|}{V (m\,s$^{-1}$)  prediction-RMSE} \\
\hline
OAM index & OAM spectrum & Intensity profile & Scintillation index & OAM spectrum & Intensity profile & Scintillation index\\
\hline
3 & 0.492 & 0.536 & 1.779 & 0.029 & 0.041 & 0.167\\
\hline
2 & 0.524 & 0.671 & 1.833 & 0.035 & 0.053 & 0.173\\
\hline
1 & 0.474 & 0.730 & 2.005 & 0.037 & 0.046 & 0.153\\
\hline
0 & 0.694 & 0.647 & 1.796 & 0.061 & 0.076 & 0.191\\
\hline
-1 & 0.503 & 0.494 & 2.048 & 0.026 & 0.039 & 0.146\\
\hline
-2 & 0.645 & 0.688 & 2.109 & 0.048 & 0.076 & 0.177\\
\hline
-3 & 0.539 & 0.584 & 1.924 & 0.051 & 0.064 & 0.216\\
\hline
\end{tabular}
\caption{Prediction accuracy with OAM spectrum, intensity profile and scintillation index.}
\label{tab:Comparison}
\end{table}

The interaction of structured light with the environment provides multi-dimensional information about the optical path due to skewed optical rays that are inherent in beams that carry Orbital Angular Momentum (OAM). This skew is determined by the radius and OAM carried by a particular probe beam, where this angle is determined by the equation $\frac{\ell}{k\,r}$, where $k$ is the optical wavenumber and $r$ is the radius of a particular point in on the wavefront. These screwed rays result in a form of path diversity, similar to multi-aperture scintillation sensors (MASS); however, with geometrical similarity to vortices induced by turbulent flow. In our approach, we focus on bulk turbulent channels as the cascaded interactions with turbulence impart structural changes to the optical field in 3D, which has been shown in previous work to produce unique path dependent aberrations for beams that carry OAM. The presented experimental and theoretical investigations show a distinctive advantage for the combination of structured field and machine learning for environmental sensing.

Potential application areas for this approach are the direct measurement of environmental parameters of clean air and low humidity turbulence conditions, where particulate tracking and water-vapour absorption based technologies such as differential absorption lidar (DIAL) techniques are not applicable. The presented approach requires a transmitter and receiver architecture, and not back reflection, which means it is not directly competitive with established technologies such as DIAL, RADAR or LIDAR. As spatial information is required for effective measurement of structured light, the approach will be bound by the spatial bandwidth of the optical system. This spatial bandwidth can be estimated by considering the Fresnel number of any given channel. This is commonly defined as $F=a^2/d\lambda$, where $a$ is the aperture size at the receiver and $d$ is the length of the optical channel. For receiver apertures of 100\,mm and optical probe beams with $\ell=3$, the expected operational range is around 1\,km for 850\,nm light. Sub-km weather modeling is an emerging field that is anticipated to assist in the prediction of extreme weather events \cite{Smith2021Jan}. Novel technologies could improve the accuracy of path integrated weather sensing as sub-km scales could be beneficial for developing accurate models. Our approach was demonstrated over 36\,m, but could readily be applied at shorter distances of several meters or up to 1\,km when appropriate aperture sizes are used. One area of application could be for compact thermal and airflow measurement of the exhaust from jet engines to assist in performance monitoring and design of more efficient engines \cite{Petzold1998Jan, Wang2017Jan}. 

SDM approaches, such as MIMO and OAM multiplexing, when deployed in free-space commonly need to actively mitigate the effects of atmospheric turbulence. Mitigation methods based on digital signal processing have received considerable interest over the last decade, where these technologies are constantly measuring the inter-channel crosstalk to determine the transmission matrix of the channel. Our approach could be used in tandem with these digital signal processing approaches to measure environmental parameters both in general sensing for weather monitoring or additionally for predicting potential channel errors that could occur to support communication control and automation strategies. Fading models can be used to identify expected statistics on expected deep fades and crosstalk that can be used for selecting the proper error correction or fading resiliency schemes \cite{Majumdar2008, Khalighi2014Jun}.

In conclusion, we have demonstrated the use of OAM modes propagation in turbulent environments as a probe for accurately measuring windspeed and temperature. We have demonstrated that the use of OAM mode decomposition, when combined with a machine learning approach that reveals trainable features in the distortions of vortex beams to allow for effective environmental monitoring. It was shown that SVM regression models to measure temperature variations of $0.49^\circ$C and wind speed variations of 0.029\,m\,s$^{-1}$  controlled free-space channels with short 3\,s measurements. The predictable nature of these findings could indicate the presence of an underlying physical relationship between environmental conditions that lead to specific eddy formation and the OAM spiral spectra. More generally, mode de-multiplexing combined with machine learning is a potentially powerful technique that could be deployed for a wide range of applications, where the constant evolution in relative phase between modes can be considered as noise within the system. These applications could include fiber sensing or retrieving information from dense scattering environments. Similar mode-demultiplexing approaches could be used to analyse statistical variations in relative phase to further increase system sensitivity. 

\section{Methods}
\label{sec:Methods}
\subsection{Environmental Emulator}

Key to the development of any reliable sensor technology is the availability of appropriate controllable test facilities to generate known environmental conditions. To replicate atmospheric interactions, we constructed a turbulence emulation chamber with computer controlled temperature and wind speed. Warm air is heated using electrical heaters, embedded in the thermal mass of igneous rock, where the temperature of the air can be varied to approximately $15^\circ$C above ambient room temperature. Digital thermometers were utilised to control a feedback loop to control the heating of the warm air chamber at the target temperature. As mixing of hot and cold air is required, cooler air of approximately $21^\circ$C was mechanically mixed by 30 fans over the 2\,m length of the chamber. The speed of fans can specifically be chosen from 0 to 4.95\,m\,s$^{-1}$ , where each is controlled by a computer using pulse width modulation (PWM). The fan speed is monitored by a tachometer, where closed-loop control is implemented on a custom control board, using a series of multi-channel fan controllers (Maxim Integrated MAX31790), and was interfaced to the camera by Labview over a serial USB connection. To emulate the multiple interactions with turbulence expected in long distance optical propagation, folded mirror arrangement was used to achieve a total propagation length of 36\,m. Each reflection is offset, to experience a different effective turbulence induced phase aberration. 

\subsection{Optical System}
Our optical beam preparation was achieved by illuminating a Hamamatsu LCoS Spatial Light Modulator (SLM) with an 850\,nm collimated Gaussian beam. The optical beam is encoded with a spatial structure, through the use of a blazed diffraction grating to shape both the phase and intensity simultaneously in the first order of the diffracted beam. To fully calibrate the spatial light modulator and minimize the effect of the aberrations from optical components in the system, an iterative optimisation process was used as outlined in \cite{Jesacher2007Apr}. 
The folded beam path employs 25\,mm square aperture dielectric mirrors (Thorlabs BBSQ1-E03), resulting in a transmission loss of approximately 7.65\,\% after undergoing 18 reflections. After propagation through the turbulent environment, a beam splitter was used to allow for simultaneously recording the optical beams intensity profile and to analyze the distribution of OAM modes induced by the turbulence. We utilise a passive optical device designed for the sorting of beams that carry OAM to perform a modal decomposition of the light emanating from the turbulence box. The sorter utilized in our experiment comprised two free-form optical surfaces to perform the required transformation, where the optical profiles for each of the surfaces are fully outlined in \cite{Lavery:2012we}. The surfaces used had an aperture diameter of 12.5\,mm, which were directly machined on each side of a 10\,cm long bar of polymethyl methacrylate (PMMA), utilising the same machining method as outlined in \cite{Lavery:2012we}. This device performs a log-polar transformation, which subsequently allows a spherical lens to convert beams that carry angular momentum into a series of separated spots at the back focal plane of a lens \cite{Berkhout:2010cba}. 
A camera is placed at the focal plane of this lens, where the pixels in 51 adjacent, equally sized regions were summed. Each region is measuring the power associated of a specific OAM beam, over the range of modes $\ell \pm 25$. This type of measurement is often referred to as an OAM spectrum to draw an analogy to optical frequency spectroscopy. 

\begin{figure}
   \center
   \includegraphics[width=0.75\textwidth]{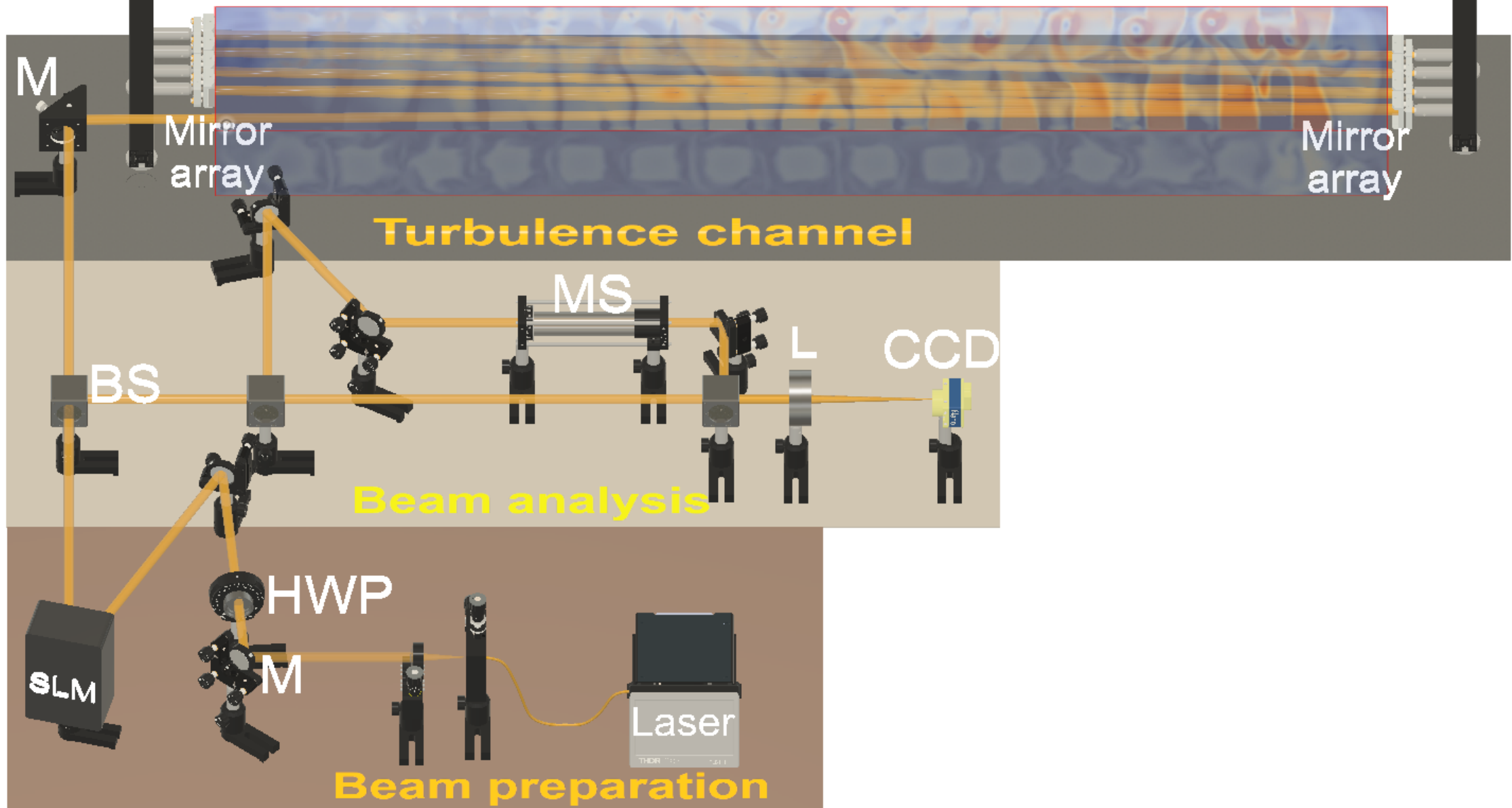}
   \caption{Experimental setup. The setup is divided into three parts: beam preparation, turbulence channel, and beam analysis. M, mirror, HWP, for half wave plate, SLM, for spatial light modulator, BS, beam splitter, MS, mode sorter, and L, Lens.}
   \label{fig:ExptSetup}
\end{figure}

\subsection{Acknowledgements}
The authors would like to thank  Manlio Tassieri (Uni. of Glasgow) for discussions on fluid dynamics.  We thank Lucia Uranga for proofreading services. The work was supported by the Horizon 2020 Future and Emerging Technologies Open grant agreement "Super-pixels" No. 829116 and the EPSRC grant EP/T009047/1.

\section{Supplementary}

\subsection{Greenwood frequency}
The Greenwood frequency represents the necessary frequency or bandwidth for achieving the best correction with an adaptive optics system. It can be calculated with the equation (\ref{eq:fG}) \cite{Fried1990May,Yao2019Sep}

\begin{equation}\label{eq:fG}
    f_G = 2.31 \lambda^{-\frac{6}{5}} \left[ \sec \theta \int_{\text{channel}}  C_\mathrm{n}^2(z) V(z)^{\frac{5}{3}} dz \right]^{\frac{3}{5}},
\end{equation}
where $\lambda$ represents the wavelength, $\theta$ signifies the zenith angle, $V(z)$ stands for the windspeed, and $C_{n}^{2}(z)$ denotes the turbulence constant structure function. In our computation, $\theta$ is assumed to be 0. The turbulence constant structure function $C_{n}^{2}(z)$ is determined through the scintillation of total energy at the receiving plane \cite{Andrews2019}. The scintillation index is computed over a duration exceeding 500\,s, employing a 3\,s averaging time window.

\begin{table}[H]
\centering
\begin{tabular}{|c|c|c|c|c|}
\hline
\multicolumn{5}{|c|}{Greenwood frequency} \\
\hline
RPM & V (m\,s$^{-1}$) & $f_{G_{\text{min}}}$ & $f_{G_{\text{max}}}$ & $f_{G_{\text{mean}}}$\\
\hline
1000 & 1.34 & 21.51 & 126.78 & 42.34\\
\hline
1100 & 1.42 & 26.32 & 289.90 & 72.77\\
\hline
1200 & 1.58 & 28.30 & 222.15 & 61.31\\
\hline
1300 & 1.73 & 20.90 & 175.15 & 56.49\\
\hline
1400 & 1.89 & 28.28 & 100.63 & 67.59\\
\hline
1500 & 2.23 & 47.28 & 120.22 & 87.29\\
\hline
1600 & 2.35 & 52.68 & 388.12 & 92.44\\
\hline
1700 & 2.52 & 31.31 & 381.95 & 84.60\\
\hline
1800 & 2.67 & 22.63 & 253.91 & 69.08\\
\hline
1900 & 2.81 & 54.0 & 376.39 & 91.93\\
\hline
2000 & 2.92 & 56.96 & 294.42 & 95.63\\
\hline
\end{tabular}
\caption{Measured Greenwood frequency. The input OAM index is 3 and the average time is 3s.}
\label{tab:Greenwoodfrequency}
\end{table}

\begin{figure}
   \center
   \includegraphics[width=0.8\textwidth]{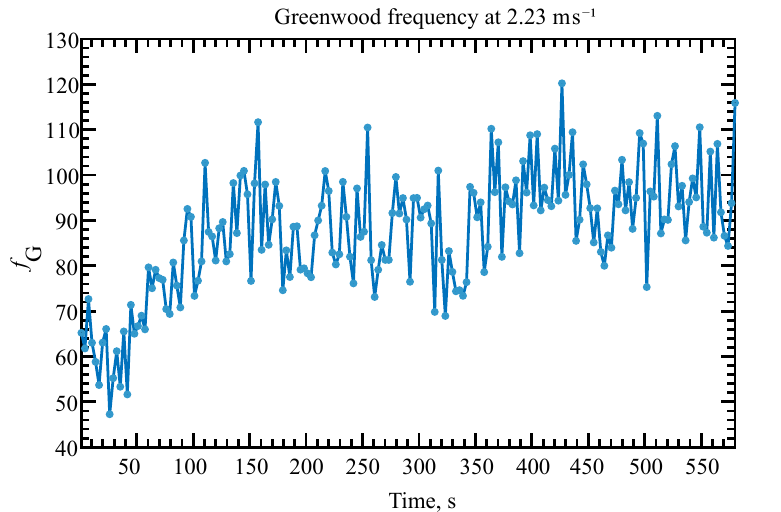}
   \caption{Measured Greenwood frequency. The input OAM index is 3 and the average time is 3s. The wind speed is 2.23\,m\,s$^{-1}$ .}
   \label{fig:ExptSetup}
\end{figure}

\end{document}